# Intentional Design for Empowerment

Noah S. Podolefsky

*Department of Physics, University of Colorado Boulder, 390 UCB, Boulder, CO 80309*

**Abstract:** I argue for empowering education, adapting Marx's idea of ownership of the means of production, and discuss interactive simulations as one example of a tool in which intentional design can support student ownership of learning. I propose a model that leverages affordances of educational tools to do positive work toward empowering education.

## INTRODUCTION

Jane sits at a computer using an interactive simulation (sim) during a think aloud interview. She has been told to use the simulation in any way she chooses. The simulation is Circuit Construction Kit (CCK), a college-level sim designed by the PhET project that allows students to build circuits with batteries, bulbs, and wires. Jane is in 4$^{th}$ grade. Within a few minutes, Jane has dragged out a bulb, battery, and wire and set a challenge for herself: get the bulb to light. She spends the next 30 minutes trying to complete this challenge, trying batteries in different places, using a switch, and rearranging components. Along the way, the interviewer discusses some of Jane's ideas, asking questions but never telling her what to do. Around 30 minutes, she has an idea – the blue dots (electrons) need to move, and they can't move if they don't have anywhere to go. She realizes she needs to give the electrons a path, and drags out wires to complete the circuit. She makes the circuit, but it does not work – the switch is open. She closes the switch, the bulb turns on, and Jane throws her hands in the air, triumphantly exclaiming "Yes! Yes!"

Jane's experience is not unique. In interviews with elementary, middle school, and college, I often observe students taking great joy setting their own challenges and exploring science using PhET sims. These actions demonstrate a high level of autonomy and ownership over PhET sims, which I have also observed in classrooms. For instance, at the end of one class using a sim-based activity, a student asked "Can we keep doing this instead of going to recess?"

In this paper, I focus on student autonomy and ownership over their learning experiences. I argue for the importance of ownership over how knowledge is produced, and suggest that as researchers and educators we must be intentional about design decisions that affect students' perception of who has agency in learning activities. Throughout this paper, I take the stance that the main purpose of education is individual and collective social empowerment. All elements of education – lecture, labs, simulations, textbooks, etc. – serve this goal. I follow the traditions of Dewey, Arons, and others to examine our practices with a political lens, toward a better society.

## PLAY AND OWNERSHIP

As part of conducting research with the PhET group, I have been collaborating with K12 teachers in order to create and implement sim-based activities. A standard practice in these activities is to include 5-10 minutes of open play before students begin a more guided activity. During open play, students use a PhET sim without instructions. I have repeatedly observed students asking questions and sharing ideas with other students. Students will get up from their desks and go to another group to discuss their ideas. Ideas "go viral" throughout the classroom. These are demonstrations of both individual and collective empowerment, where students take ownership over the learning tools, and the experiences, in which they participate in classrooms.

Without this open play time, classroom discourse and facilitation by the teacher can be markedly different. One study compared a class where students had 5 min of open play time with a sim to a class without open play time, and found that the teacher had significantly more difficulty leading a student-centered, inquiry activity when no open play time was allowed.[1] Thus, student control over the learning process not only instills a sense of ownership for students, but can also support teachers to carry out classroom discourse in which the students are active participants in the learning process.

The design of the learning environment and tools can support teachers and students to work together in learning about science content, and feel themselves to be cooperative partners in this enterprise.

## OWNERSHIP OF THE MEANS OF PRODUCTION

In order to understand the changing power dynamic described above, I borrow from Marx's concept of "ownership of the means of production".[2] In discussing labor, Marx argued that it was essential for workers to own the means of production of material goods. When this ownership is taken from workers and transferred to business owners, workers become alienated from both the products of their labor and from the meaningfulness of their labor itself. Furthermore, under this system of alienated labor, workers become alienated from each other, and competition, rather than cooperation, becomes the main way of "getting ahead".

Schwartz[3] has pointed out a connection between alienation felt by workers and alienation felt by students. Lave and McDermott[4] analyzed Marx's *Estranged Labor*, replacing *Labor* with *Learning*. They argue that like workers described by Marx, when *learners* lose ownership over the means of production of *knowledge*, they become alienated from both the product and process of learning. Further, they enter into a relationship that is competitive, rather than cooperative, between themselves, teachers, and other participants in the school system.

I suggest that many school activities guide student learning in ways that may be productive for content learning, but may take ownership from students. The challenge I set out is how to achieve some goals that are established norms of schooling (e.g. content goals), while returning ownership of the means of production to students and creating a more empowering situation for all participants in the education system.

## MENO'S PARADOX

Establishing ownership of the means of production is not as simple as handing over the reins to students without further reflection. This has been understood for over two millennia. In Meno, Plato describes the paradox: if a person knows something, then they do not have to learn it. However, if they do not know something, then in order to learn it they need to know what it is they do not know, and how to go about learning it. Thus, left to their own devices, people cannot come up with answers on their own, nor can they come up with the *questions*.

The way this paradox is often solved in schooling is to make schooling *normative*. So called "old timers" know the questions and the answers, and they instruct and enculturate "new comers" by enacting the curriculum through instruction. This instruction can be didactic, where the teacher holds the authority for choosing the questions and providing the answers. With inquiry approaches, the responsibility for answer making, or in some cases sense making, is placed on students. However, which questions are asked, and what the answers ultimately should be, is set by the curriculum. These approaches may limit the degree of ownership realized by students (and teachers).

## IMPLICIT SCAFFOLDING

The issues I describe above are considerable, with great momentum. Here, I propose that two tools can help address these issues – an intellectual tool, the implicit scaffolding framework, and a material tool, interactive simulations. I do not claim that these tools alone can create sweeping change, but discuss them as exemplars of empowering tools for change. Critical to my argument is that the design of these tools is *intentional* toward certain goals, such as ownership.

Implicit scaffolding is based on a constructivist model of learning, design principles from human computer interaction[5], and over 10 years of research on development and implementation of PhET sims.[6,7] Constructivism states that learners must be actively engaged in the learning process. This defines a fundamental cognitive process, but underspecifies the pedagogical practices that support constructing knowledge. Constructivism applies even to learning from pure lecture – if students do learn in this environment, it is because they are actively engaged in listening, and not simply "absorbing" knowledge. At the other end of the spectrum, students could be engaged in "pure discovery" with little to no structure to guide their learning. Somewhere between these extremes is guided inquiry, where student learning is scaffolded through prompts, questions, or instructions.

A typical example of a guided activity is the introductory physics lab. These consist of a set of instructions for carrying out an experiment, with data collection and a set of questions for students to answer. These activities leave little room for ownership on the part of students, with nearly all of their actions guided explicitly by the lab instructions. More student-centered, inquiry-based activities might have more leading questions than specific instructions, or ask students to do sense making or design an experiment. Even these milder forms of guidance may still leave students with the task of question answering, rather than question asking. I assert that *question asking* is a critical component of ownership, and thus I seek pedagogical tools that enable question asking by students. I would like to design tools that allow students to guide their own learning, while circumventing Meno's paradox.

The basic premise of implicit scaffolding can be exemplified by examining the doors shown in Figure 1. Both doors need to be *pushed*. The door on the left has a handle which is a strong cue that the door should be *pulled*. The owner of this door attempted to fix this design problem by writing "PUSH" on the door. However, as reported by the door owner, this did not keep him from often pulling the door even after years of using it. The door on the right has no handle. Both doors *afford pushing*, but the door on the right is *constrained* so that it cannot be pulled. Through use of affordances and constraints, correct usage of the door is *guaranteed by design*. No instruction is necessary, allowing users to get past the frustration of using the door incorrectly and get on with leaving the building.

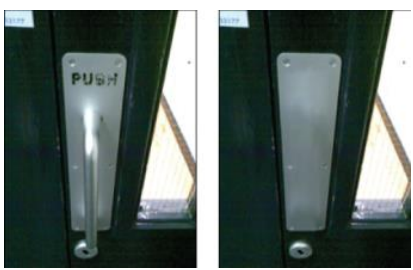

**FIGURE 1.** Poorly- and well-designed doors.[8]

Similarly, interactive sims can afford actions that are productive for learning, and constrain actions that might be confusing, frustrating, or counter-productive. For instance, in the PhET sim Energy Skate Park, students can build custom tracks of any shape and size. This freedom to build can lead to students building elaborate tracks that do not always illuminate key learning goals (i.e., conservation of energy). The sim was redesigned to use pre-set tracks that focus students on key energy ideas. In addition to affordances and constraints, cues are included to focus students' attention on the exchange of kinetic and potential energy, representations are used that students can make sense of without explicit instruction, and feedback allows students to create experiments and see the effects of changing variables. Track building, which can support student ownership, is retained in another part of the sim. This design approach allows students to ask their own questions and make sense of science ideas through exploration.[7] In short, implicit scaffolding *guides without students feeling guided.*

## ON BEING INTENTIONAL

I have argued that in designing curricular materials and learning environments for school – whether interactive simulations, labs, homeworks, etc. - we should be intentional about our choices, especially in terms of how those tools empower students and teachers to work cooperatively. I acknowledge that with all of the burdens and responsibilities placed on students and teachers, trying to attend to these broad social goals may lead to feeling like Atlas carrying the weight of the world on his shoulders. One might feel that although these issues are important, they are too big and burdensome to be tackled all at once. I agree, and suggest that people take the common credo to "think globally and act locally." While this lessens the burden, it still requires participants in the system to act.

What happens if we do not act with intention? The answer is that we cannot "do no harm" if we are not vigilant and intentional about how the normative activities of schooling affect socio-political issues such as ownership of the means of production. It has been said that "the things you own end up owning you."[9] Tools that we create for teaching and evaluating students can take on a mind and role of their own, with unintended socio-political consequences.

I deal with this very issue as a designer of interactive sims. Whereas sims can be used for students to explore in a way that is empowering, these tools can and often are used in very traditional ways, such as for didactic lecture or cookbook labs. Neither of these uses have the effect of providing students ownership over their learning. Simultaneously, we should not take ownership from the *teachers* who find these methods useful for their teaching, so we are caught in a conflict, in some sense subject to the personalities our creations have taken on and the socio-political issues they entail. It is not yet clear how to resolve this conflict in a way that empowers all participants in the school system.

Consider another example from the Force Concept Inventory (FCI).[10] One of the questions involves a canon on a cliff, and asks students to select the path the cannon ball would take when fired. There are 5 answer options, and to a physics expert the correct answer is clear. However, an analysis of this question using the distances show in the problem figure results in an initial speed of the canon ball of about 13 m/s. A real canon ball travels on the order of 300 m/s. From this analysis, it is clear that the *correct* answer is not one of the choices (the ball should move in a nearly straight line off of the right side of the page). Mazur[11] has pointed out similar issues with some concept questions.

Why point out this discrepancy? In my view, in answering this question there is much more going on than students demonstrating their knowledge of Newton's laws. They are also demonstrating a habit of mind, learned through schooling, that when questions are asked, there are allowed ways of interpreting those questions, and thus allowed answers. Answering this question on the FCI is as much a matter of epistemic framing[12] as it is conceptual understanding. In fact, in answering with this epistemic framing, a student should ignore all of the realities of how real canons work (including the fact that they are used in warfare) and answer the very narrowly defined question of which answer choice to select. Most physicists would probably think that this sort of answer making is not reflective of critical thinking, and in fact is indicative of an unfortunate narrowing of what students notice when examining a physical situation. I agree – my point here is that a tool (the FCI) that was created with the best of intentions (measuring conceptual understanding) may reflect a narrowing of students' critical thinking and noticing of inconsistencies in the situations they are presented with.

## DISCUSSION

Some have argued that contrary to the conventional wisdom that the purpose of schooling is to empower students with knowledge, one function of the school system is to disempower students by diminishing their ability and inclination to think critically in a broad sense. Chomsky[13] argues that through K12 education, children are indoctrinated into a system where they accept truths as handed down from authority, and taught to accept socially inequitable conditions as the norm. This is reinforced through grades that stratify students, and practices such as pledging allegiance to the flag that generate an empty patriotism, rather than a rich sense of community. Schmidt[14] argues that one function of higher education is to indoctrinate students into a system where they carry out their work without questioning the broader social and political consequences of that work. He claims this function serves to, for example, support scientists who may be uncomfortable with war to carry out their work designing weapons. Critical thinking is required only to the degree that it improves the products. It is not a means to individual or collective social empowerment.

It is not clear that these functions of education were designed purposefully. They can evolve within society and we need not assume a "creator". Historically, one can identify episodes of significant change to education that were intended to have social benefit but, long term, may have resulted in social stratification and marginalization, such as the shift to vocational education in the early 20$^{th}$ century, meant to increase worker productivity, but at the expense of worker empowerment. Notably, this move had its dissenters at the time,[15] and others have continued this legacy.

It should be strongly emphasized that this is not an argument against schooling as a whole. Rather, it is an argument against aspects of schooling that are disempowering, and *for* aspects that are (or could be) empowering. Students do learn useful things in school, and the school system is essential to the infrastructure that creates and sustains culture. My call is to be conscious and intentional about the culture created.

My objective here is critical in the sense that I aim for positive change, pointing out the problematic structures that exist so that we can do something about it. Thus my call for researchers and educators to be *conscious* and *intentional* about their choices and the tools they use in teaching students. I describe interactive sims as one tool which can be leveraged to create environments in which students can take ownership of their learning and realize empowerment through exploring questions that are meaningful *to them*. Implicit scaffolding is an intellectual tool that can leverage the affordances (and constraints) of tools to generate individual and collective social empowerment. Critical to these tools being effective for change is that we as researchers and educators examine and reexamine the tools we use and make sure they are still doing work for us, and not that we are now working for them.

## CONCLUSION

I have unabashedly proposed that the purpose of education is empowerment, and that some aspects of the education system may work against this goal. I go beyond critique to propose solutions, using Marx's idea of *ownership of the means of production* to consider how educational tools can work for or against empowerment, and that doing positive work requires us to be conscious, intentional, and vigilant. This is in some sense a call to arms, but not a call for war. A rising tide raises all boats – in this case, empowering the marginalized among us raises all of our humanity.

## ACKNOWLEDGMENTS

I thank members of the PER group at Colorado and the broader PER community for thoughtful discussions.

## REFERENCES

1. Podolefsky et al., PERC Proceedings (2012)
2. K. Marx, *Estranged Labor* (1844)
3. D. Schwartz, in Dillenbourg, P. (ed) *Collaborative learning*. NY: Elsevier Science/Permagon. (1999)
4. J. Lave & R. McDermott, *Cricital Practice Studies* (2002)
5. D. Norman, *Design of Everyday Things*, Basic Books 2002
6. A. Paul et al., PERC Proceedings (2012)
7. N. Podolefsky et al., arXiv:1306.6544
8. C. Walsh (2007) http://iqcontent.com/blog/2007/01/the-usability-of-garda-doors/
9. Palahniuk, Chuck. *Fight Club*. WW Norton, 2005.
10. D. Hestenes et al. *The physics teacher* 30 (1992): 141
11. E. Mazur, AAPT Millikan Lecture (2008)
12. A. Elby & D. Hammer. Personal epistemology in the classroom. (2010): 409-434.
13. N. Chomsky. Rowman & Littlefield Publishers, (2004)
14. J. Schmidt. Rowman & Littlefield Pub Inc., (2000)
15. D.F. Labaree, in D. Trohler, T. Schlag, F. Osterwalder (eds.) *Pragmatism and Modernities*, 163-188 (20010)